
\magnification = 1200
\hsize = 16 true cm
\vsize = 23 true cm
\global\hoffset= 0.2 true cm
\global\voffset= -0.42 true cm

\font\twelverm=cmr10 scaled\magstep1
\font\fourteenrm=cmr10 scaled\magstep2

 \phantom{.}
\noindent
To be published in the Proceedings of ISQM-SAT 92, {\it Quantum
Control \& Measuremnt}, H. Ezawa and Y. Murayama eds., North Holland,
1993.

\noindent
TAUP 2020-93, hep-th/9305001
 \vskip 1.2 cm

\noindent
{\bf {\fourteenrm  The Schr\"odinger wave is
 observable after all!}}
\vskip 0.6cm
\noindent
{\twelverm{Yakir Aharonov
and Lev Vaidman}
\vskip 0.6cm
\baselineskip 12pt plus 0pt  minus 0pt
\tenrm
\noindent
School of
Physics and Astronomy \hfill \break
Raymond and Beverly Sackler Faculty of Exact Sciences \hfill \break
Tel-Aviv University, Tel-Aviv, 69978 ISRAEL
\vskip 0.15 cm
\noindent
and
\vskip 0.2cm
\noindent
Physics Department, University of South Carolina \hfill \break
Columbia, South Carolina 29208, U.S.A.
\vskip 0.8 cm
\noindent
{\bf Abstract}
\vskip 0.05 cm
It is shown that it is possible to measure the Schr\"odinger wave of
a single quantum system. This provides a strong argument for
associating physical reality with a quantum state of a single system
in sharp contrast with the usual approach in which the physical
meaning of a quantum state is related only to an ensemble of identical
systems. An apparent paradox between measurability of a quantum state
of a single system and the relativistic causality is resolved.
\vskip 0.8cm
\noindent
{\bf 1. INTRODUCTION}
\vskip 0.32cm
The Schr\"odinger wave is the most important tool in quantum theory.
However, when
one tries to associate a physical reality with the Schr\"odinger wave of a
single particle one runs into serious difficulties, and it is
generally
believed that this task is impossible.  Many accept the
interpretation
in which  the Schr\"odinger wave is only a mathematical tool for
calculating probabilities of
(real) experiments.
Let us list some of the difficulties
with identifying the Schr\"odinger wave as physical reality.

 i) We have never seen a quantum state of a single particle in a
laboratory.
Although the wave is often spread over a region of space, we never
see a particle simultaneously in several distinct locations.

ii)  There is an argument that we will also never see the
Schr\"odinger wave in the laboratory.  Assume that we can ``see" the
quantum state.  Then, it seems, we can distinguish among different
states.  But the possibility of distinguishing between nonorthogonal
states contradicts the unitarity of quantum theory.  Indeed, the
scalar products between any pair of quantum states do not change
during unitary time evolution.  But the alleged measuring procedure
changes this scalar product to zero.

iii)  The collapse of the wave during the measurement contradicts
 Lorentz covari-\break ance [1].

If, however, the Schr\"odinger wave represents only a mathematical
tool, no such difficulties arise: we should not ``see" mathematical
objects in a laboratory and there are no limits on the rate of change
of a mathematical function.

In this letter we will show that the density $\Psi^* \Psi$ and the
current ${{i\hbar}\over{2m}}(\Psi \nabla \Psi^* -\Psi^* \nabla \Psi)$
of the Schr\"odinger wave are observable even for a
single particle and, therefore, do represent physical reality.
The usual
quantum measurements referred to in (i) alter the Schr\"odinger
wave, and therefore they are not adequate for our purpose.
We will
 use  special
{\it protective}  measurements.
Protective measurements allow us to measure the density and the
current of the Schr\"odinger wave without changing it.  In some cases
 energy conservation provides protection for the state,
 while in other cases we have to add a special
protection procedure.

The second argument is certainly correct in the claim that unitarity
prevents us from distinguishing among nonorthogonal states, but it
only implies that there is no {\it universal} procedure for observing
different states; it allows, however, for the possibility that for any
state there is an appropriate measuring procedure.  In this work we
shall show that this is indeed so.

The last problem is the most serious one.  Assume that we start with a
particle in a superposition of being in two separate boxes, and then
find it in one of them: How did half of the wave moved instantaneously
from one box to another?  Recently, we have developed a novel approach
which helps to solve this difficulty [2].  Here, however, we will only
touch this issue by considering the apparent paradox between the
causality requirement and our Schr\"odinger wave measurements.  Our
solution also yields an answer to the recent proposal for superluminal
communication [3].
\vskip 0.8cm
\noindent
{\bf 2. MEASUREMENTS OF NONDEGERNERATE EIGENSTATES}
\vskip 0.32cm
Let us start by discussing the simple case in which the system itself
supplies the protection of the state and no artificial protection is
needed.  This is the case of a discrete nondegenerate eigenstate of
energy.  Now, every sufficiently slow and weak measurement, i.e., any
measurement with adiabatic interaction, will not destroy the state of
the particle.  And, if the measurement is long enough, it can provide
any desired precision.

The first example is a measurement of the  ground
state of the electron in the hydrogen atom.  The state
$\Psi(x)$ can be
chosen real and positive aside from the trivial time dependent part.
Therefore, it is enough to measure the expectation values of
projection operators $P_{V_n}$ on small regions $V_n$ where

\vskip 0.6cm
\noindent
 $
P_{V_n}(x)
 \equiv \cases{
1,&if $x \in V_n$;\cr 0,&if $x \not\in V_n$~.\cr}
\phantom{{-i}\over{2}}
\hfill (1) $

\vskip 0.6cm
\noindent
The expectation value of $P_{V_n}$ yields the value of $|\Psi|^2$
times the volume of the region $V_n$, and since $\Psi$ is not negative
we obtain $\Psi$ everywhere in space.

In order to measure $P_{V_n}$ we will use the standard von Neumann
measuring procedure, but instead of an instantaneous interaction we
will make it long and adiabatic. The interaction Hamiltonian will be

\vskip 0.6cm
\noindent
$
 H
 = g(t) p P_{V_n}, \phantom{{-i}\over{2}} \hfill
\phantom{{-i}\over{2}} (2)
$
\vskip 0.6cm
\noindent
with $g(t) = 1/T$ for a period of time $T$ and smoothly going to zero
before and after.  The initial state of the measuring device will be
chosen so that the canonical conjugate $p$ of the pointer variable $q$
will be effectively bounded (say, a Gaussian centered at zero).  For
$g(t)$ smooth enough we obtain an adiabatic limit in which the
electron cannot jump from one energy eigenstate to another.  For
bounded $p$, in the limit of $T \rightarrow \infty$, the interaction
Hamiltonian goes to zero.  The energy of the eigenstate shifts by an
infinitesimal amount given by first-order perturbation theory:
\vskip 0.6cm
\noindent
$
\delta E =\langle H_{int}\rangle   =
{{\langle P_{V_n}\rangle p}\over T}. \hfill (3)
$
\vskip 0.6cm
\noindent
In the limit $T \rightarrow \infty$ the energy eigenstate remains
unchanged.  Although the interaction vanishes in this limit, its
duration increases accordingly such that the measuring device yields
the desired result.

\vskip 0.6cm
\noindent
$
e^{-{i\over \hbar}p \langle P_{V_n}\rangle}.\hfill
\phantom{{-i}\over{2}} (4)
$

\vskip 0.6cm
\noindent
Thus, the state of the pointer position $q$ is shifted during the
interaction by the expectation value $\langle P_{V_n}
\rangle
$.  This is a
measurement of $\langle P_{V_n}
\rangle
$ with a precision equal to $\Delta q$,
the width of the initial pointer variable distribution. (Note that
$\Delta q$ can not be infinitesimally small because
 $\Delta p$ is bounded.)

Obviously, this procedure will also work if we couple the electron to
many such measuring devices, thus measuring the projection
operators in different regions, namely,

\vskip 0.6cm
\noindent
$
 H = \sum g(t) p_n P_{V_n}.\phantom{{-i}\over{2}} \hfill
\phantom{{-i}\over{2}} (5)
$

\vskip 0.6cm
\noindent
Performing these measurements simultaneously will require a longer
measurement period but,
in principle, we can measure the value of $\Psi$ of a
single particle everywhere in space.

Protection due to energy conservation also suffices for measurement of
stationary nondegenerate eigenstates with nonzero current.  These are,
for example, the eigenstates of particles described by a Hamiltonian
 with a vector potential, such
as electrons confined to a toroid through which there is a flux of
magnetic field (the Aharonov-Bohm effect).  In this case the state
cannot be chosen real and we have to find its local phase.  (The
overall phase is unmeasurable and has no physical meaning.)  To this
end we will measure, in a similar adiabatic way, the following
operators:

\vskip 0.6cm
\noindent
$
B_n ={{-i}\over{2}} (P_{V_n}\nabla + \nabla P_{V_n}).\hfill (6)
$

\vskip 0.6cm
\noindent
The expectation values of these operators are proportional to
the current
density at these regions, and using also the
values of the density
itself we may find the gradient of the phase.  Indeed, if we represent
the Schr\"odinger wave as $\Psi (x) = r(x) e^{i\theta(x)}$ then

\vskip 0.6cm
\noindent
$
{{\langle B_n
\rangle
} \over {\langle P_{V_n}\rangle}} = \nabla \theta .\hfill (7)
$

\vskip 0.6cm
\noindent
The phase can be found by integration of its gradient  up to an
(unobservable)
constant. Note, that if there are regions which are disconnected one
from the other, then, by this method, the relative phase  cannot be
defined.

The density $|\Psi|^2$ multiplied by the charge of the electron yields
the effective charge density.  An adiabatic measurement of the Gauss
flux out of a certain region must yield the expectation value of the
charge inside this region.  Similarly, in the case of a stationary
state with non-zero current, an adiabatic measurement of the Ampere
contour integral around a small loop yields the current density times
the charge of the electron.

\vskip 0.8cm
\noindent
{\bf 3. THE OUTCOMES OF THE ADIABATIC MEASUEREMENTS ARE \phantom{3. }
SINGLE-TIME
PROPERTIES OF A QUANTUM SYSTEM}
\vskip 0.32cm
We have shown the observability of a quantum stationary state.  This
is our main argument for associating physical reality with a quantum
state.  Since our measurement lasts a long period of time we do not
have a method for measuring the Schr\"odinger wave at a given time.
Thus, we have a direct argument for associating physical reality with
stationary Schr\"odinger waves only over a {\it period} of time.  The
reader may suspect that what we have measured represents time-averaged
physical properties of the system.  Let us present a few arguments
explaining why we should associate the outcomes of our measurement
with single-time properties of our quantum system.

The essential feature of our adiabatic measurement is that the state
$|\Psi \rangle$ does not change throughout the experiment.  Since in
the standard approach to quantum theory the Schr\"odinger wave yields
the complete description of the system, we conclude that the action of
the system on the measuring device is the same at any moment during
the measurement.

However, perhaps there is a description of a system beyond its quantum
state which does change during the measurement process, and the
Schr\"odinger wave we measure does not have inherent meaning at a
given time but represents only a time average over the period of the
measurement.  Indeed, it is possible to construct a simple classical
model in which the outcomes of our measurements follow our
predictions, but no physical meaning can be associated with the wave
at a given time.  Consider a model of an atom in which the electron
performs very fast ergodic motion in the region corresponding to the
quantum cloud.  The charge density might be either zero (if the
electron is not there) or singular (if the electron is inside the
infinitesimally small region including the space point in question).
In spite of this fact, the measurement we have described will yield
outcomes corresponding to a nonsingular charge density cloud.  What is
measured here is the time average of the density, or how long time the
electron spent in a given place.

In order to see that this picture is inappropriate for the quantum
situation let us consider another example: a particle in a one
dimensional box, say in the first excited state.  The spatial part of
the state is $\sqrt{2/L} \sin (2\pi x/L)$.  The adiabatic measuring
procedure described above will yield   the Schr\"odinger
wave
density
  $(2/L) \sin^2 (2\pi x/L)$.  In particular, it equals zero at
the
center of the box.  If there is some kind of underlying position of
the electron which changes in time such that the result of the density
measurement is proportional to the amount of time the electron
``spends" there,
then half of the time it must be in the left half of the box and half
of the time in the right half of the box.  But it can
spend no time at the center of the box, i.e., it must move at infinite
velocity at the center. It is
absolutely unclear what this ``position" of an electron might be.
There {\it is} a theory [4] which introduces a
``position" for a particle
in addition to its Schr\"odinger wave; but according to this theory,
the ``velocity" of the particle in this energy eigenstate vanishes:
it does not move at all.

The mathematical formalism yields an additional argument: in our
measurement for any, even very short, period of time the measuring
device shifts by an amount proportional to $\langle A \rangle $, the
expectation value of the measured variable, rather than to one of the
eigenvalues. Thus, the expectation values,
which are the
mathematical characteristics of
Schr\"odinger waves, can be associated with very short periods of
time. In the limit the expectation values and, therefore, the quantum
state itself become single-time properties of a quantum system.

In order to explain this behavior of the measuring device we note
that its wave function is a superposition of states shifted by the
eigenvalues of the measured operator. These shifted waves are not
orthogonal and we obtain an interference effect.  The interference
miraculously yields a shift proportional to the expectation value.
The shift, however, is also proportional to the time interval, and for
a short time this shift is much smaller than the uncertainty, so we
cannot see it from a measurement performed on a single particle during
a short time.  Only the total shift accumulated during the whole
period of measurement is much larger than the width of the initial
distribution, and therefore, it is  observable on a single particle.

One more argument showing that our measurements reflect values
which are not of statistical character follows from consideration
of measurements performed on an ensemble of identical systems.
Simple calculations show that in the limit of a very large
ensemble and correspondingly very weak
and slow measurement we obtain knowledge of the quantum state of
these systems without exciting even one system in the
ensemble!

\vskip 0.8cm
\noindent
{\bf 4. MEASUREMENTS WITH ARTIFICIAL PROTECTION}
\vskip 0.32cm
The next question is how to measure the Schr\"odinger wave when it is
not protected by energy conservation, i.e., when it is not a discrete
nondegenerate energy eigenstate.  In fact, the measurement remains the
same, but we must add a ``protection" procedure.

  If we have a
degenerate eigenstate, then one of the simplest ways is to remove the
degeneracy by changing the energies of the other states, such that the
state to be measured remains unchanged, but is now protected by energy
conservation.

In case the state is a superposition of different energy eigenstates,
then the simplest way to protect the time-dependent Schr\"odinger wave
is by dense state-verification measurements which test and protect the
time evolution of the quantum state.  This is a Zeno type protection.
If we are interested in all details of this time dependent state we
cannot use measurements which are too slow.  Thus we need stronger
protection.  For measurement of any desired accuracy of the
Schr\"odinger wave, there is a density of the projective measurements
which will protect the state from being changed due to the measurement
interaction.  The time scale of intervals between consecutive
protections must be much smaller than the time scale of changing the
Schr\"odinger wave.  (It is better, however, not to make the
protective measurements too dense, otherwise they might force a time
evolution which has no connection to the undisturbed evolution of the
quantum system.)

The conceptual disadvantage of measurement with artificial protection
is that we have to know the state in order to arrange a proper
protection.  One might object, therefore, that we obtain no new
information from our measurement.  However, we can separate the
protection procedure and the measuring procedure: one experimentalist
provides protection and the other measures the Schr\"odinger wave
itself.  Then the second experimentalist does obtain new information.
Even for dense projective measurements, most of the time the system
evolves according to its free Hamiltonian, so we are allowed to say
that what we measure is the property of the system and not of the
protection procedure.  But the most important point is that we
actually ``see" the Schr\"odinger wave of a single particle using
a standard measuring procedure.

\vskip 0.8cm
\noindent
{\bf 5. MEASUREMENTS OF THE SCHR\"ODINGER WAVE AND SUPER-\break
\phantom{MS}LUMINAL
COMMUNICATION}
\vskip 0.32cm

One of the important features of non-relativistic quantum theory is
that despite the non-relativistic character of the ``collapse" of the
Schr\"odinger wave, one cannot use the collapse for sending signals
faster than light.  The possibility of measuring the value of a
quantum state at a given location at first seems to allow such
superluminal communication [3].  Let us show that our measurements of
Schr\"odinger waves do not violate causality.

The most naive way of sending superluminal signals is as follows.
Consider a particle in a superposition ${1\over\sqrt{2}} (|1\rangle +
|2\rangle )$ of being in two boxes separated by a very large distance.
The expectation value of the projection operator on the first box in
the initial state is $\langle P_1 \rangle =1/2$.  This must be the
outcome of the measurement performed in the first box.  If, however,
just prior to our adiabatic measurement in the first box, someone
opens and ``looks" into the second box, causing collapse into a
localized state in box 1 or in box 2, then the outcome of the
measurement of the projection operator in the first box will
drastically change: we no longer find $\langle P_1 \rangle = 1/2$ but
rather 0 or 1 (if the other observer found 1 or 0 respectively).
Therefore, it seems that by strong measurement in box 2 we can send
information to box 1 located arbitrarily far away.

The above argument is not valid for a very simple reason: the state
${1\over\sqrt {2}} (|1\rangle + |2\rangle )$ is not a discrete
nondegenerate eigenstate.  Since there is no overlap between the
states $|1\rangle$ and $|2\rangle$, the orthogonal state
${1\over\sqrt{2}} (|1\rangle - |2\rangle)$ has the same energy.  Thus,
there is no natural protection due to the energy conservation, and our
measurement requires an artificial protection.  The latter, however,
includes in this situation nonlocal, explicitly nonrelativistic
interactions.  The artificial protection is the source of the alleged
superluminal signal propagation.  (This is another disadvantage of
measurements with artificial protection.)

Much more subtle is the resolution of the apparent paradox for another
situation.  Consider a particle bound in a potential well in a ground
state
of energy $E_0$.  We perform an adiabatic
measurement of the value of $\Psi^*\Psi$, which is non-zero, somewhere
inside
the well.
 The outcome of the measurement has to
be this value, call it $\alpha$. Even if we
repeat the measurement many times we always have to obtain $\alpha$.
Suppose now that someone just prior to our measurement
performs a
strong measurement searching for the particle at a distance $l$ far
outside the well.  There is an exponentially small but non-zero
probability that he will find the particle there:

\vskip 0.5cm
\noindent
$
 prob_1 = e^{-2l \sqrt{2m|E_0|}},\hfill
\phantom{{-i}\over{2}}  (8)
$

\vskip 0.5cm
\noindent
(we take from
now on $\hbar = 1$).
Thus,
 if the measurements were performed on an ensemble of particles, a
few
of the outcomes of our adiabatic measurements have to be zero instead of
$\alpha$. (The other outcomes are affected too, but
this effect
is probably unobservable.) We have
presented a procedure which seemingly transmits superluminal
signals a distance $l$ from the distant observer to the well: if
the observer made the measurements then some of the outcomes of the
Schr\"odinger wave
measurements have to be zero, and if the observer did not
perform any measurement, all outcomes have to be $\alpha$.

The resolution  of the paradox involves  the
relativistic
formula: $E =mc^2$. It is necessary
that the minimal time for our adiabatic measurement (without any artificial
protection procedure) is larger than the time it takes light to
arrive from the location of the particle, $T \geq l/c$. However, if the
adiabatic measurement takes
time $T$, there is small but non-zero probability for it to
kick the particle out of its bound state:

\vskip 0.5cm
\noindent
$
prob_2 \geq e^{-2T|E_0|}.\hfill
\phantom{{-i}\over{2}} (9)
$

\vskip 0.5cm
\noindent
In this case the measurement will give a mistaken  result which
might be
zero.  The requirement that the probability  of error should not
be
smaller than the probability of finding the particle at the
distance $l=Tc$ from the potential well implies

\vskip 0.5cm
\noindent
$
e^{-2T|E_0|} \geq  e^{-2Tc
\sqrt{2m|E_0|}},\hfill
\phantom{{-i}\over{2}} (10)
$

\vskip 0.5cm
\noindent
and therefore $|E_0|  \leq 2mc^2$. Thus causality is
fulfilled if Eq.(10) is valid. In other words, if the binding energy of
a single particle is larger than $2mc^2$, then  there  is a paradox
associated  with causality.
This limitation is easily understood in relativistic mechanics: it
is the regime where pair production must be taken in account.

\vskip 0.7cm
\noindent
{\bf 6. CONCLUSIONS}
\vskip 0.3cm
We have shown that expectation values of quantum variables and
the quantum state
 itself have physical meaning, i.e., they are
measurable for individual quantum systems.  This is in sharp contrast
to the standard approach in which
expectation values
 and the
Schr\"odinger wave are statistical properties of ensembles of
identical systems.

Although our discussion is based on Gedanken
experiments,
recent experimental work with so-called ``weak links" in quantum
circuits shows that slow adiabatic measurements of the Schr\"odinger
wave can be performed in the laboratory [5].

Finally, we have shown that adiabatic measurements cannot be
used for
superluminal transmission of signals via collapse of the
Schr\"odinger wave.

\vskip 0.8cm
\noindent
{\bf Acknowledgements}
\vskip 0.32cm

It is a pleasure to thank Sidney Coleman, Shmuel Nussinov and Sandu
Popescu for helpful discussions.
 The research was supported by grant 425/91-1 of the the Basic
Research Foundation (administered by the Israel Academy of Sciences and
Humanities).
\vskip 0.8cm
\noindent
{\bf  References}
\vskip 0.32cm

\noindent
1~~Y. Aharonov and D. Albert, {\it Phys. Rev.} {\bf D24} (1981)
359.\hfill \break
2~~Y. Aharonov and L Vaidman, {\it Phys. Rev.} {\bf
A41} (1990) 11. \hfill  \break
3~~T.P. Spiller, T.D. Clark, R.J Prance, and H. Prance, report at
ISQM, Tokyo (1992). \hfill  \break
4~~D. Bohm, {\it Phys. Rev.} {\bf 85} (1952) 166. \hfill  \break
5~~T.P. Spiller, T.D. Clark, R.J Prance, and A. Widom, {\it
Prog. Low Temp. Phys.}  XIII \phantom{5~~}(1992) 219.

\bye